\documentclass[12pt]{article} 
\usepackage{latexsym, epsfig}
\date{6 April, 2006}
\begin{document}
\title{Thermal stability of radiant black holes\footnote{Dedicated to
the memory of Prof. A. K. Raychaudhuri}} 
\author{Parthasarathi Majumdar\footnote{email: parthasarathi.majumdar@saha.ac.in} \\Theory
Group,  Saha Institute  of Nuclear Physics, Kolkata 700 064, India.}
\maketitle
\begin{abstract}
Beginning with a brief sketch of the derivation of Hawking's theorem of
horizon area increase, based on the Raychaudhuri equation, we go on to
discuss the issue as to whether generic black holes, undergoing Hawking
radiation, can ever remain in stable thermal equilibrium with that
radiation. We derive a universal criterion for such a stability, which
relates the black hole mass and microcanonical entropy, both of which
are well-defined within the context of the Isolated Horizon, and in
principle calculable within Loop Quantum Gravity. The criterion is
argued to hold even when thermal fluctuations of electric charge are
considered, within a {\it grand} canonical ensemble. 
\end{abstract}

\section{Introduction}

\begin{quote}
... the equation for the rate of change of  expansion (Raychaudhuri's
equation) plays a central role in the  proofs of the singularity
theorems. \cite{hawel}
\end{quote}

Even in its fifty-first year, the Raychaudhuri equation \cite{akr}
retains its prime position as a tool to analyze spacetime
singularities (for a review for non-specialists, see
\cite{pm}). However, lest one should feel that the equation has only
one specialized role in general relativity, namely that of delineating
singularities, we remark that it has had far wider applications, as
for example in establishing the so-called laws of black hole
mechanics. The earliest of these laws or theorems is called Hawking's
theorem of increase of horizon area of a black hole. It states that
\cite{wald}
\begin{quote}
{\it The area of the event horizon of an isolated stationary black
hole can never decrease in any physical process}
\end{quote}

We recall first that a black hole spacetime ${\cal B}$ is the set of
events that lie in the {\it complement} of the chronological past of 
events infinitely
distant and infinitely far in the future (the future asymptotic null
infinity ${\cal I}^+$). The event
horizon of the black hole is the boundary $\partial {\cal B}$ 
of events in spacetime accessible to such observers. It is a 
null 2+1 dimensional surface hiding the black hole singularity. 

\begin{figure}
\begin{center}
\epsfxsize=70mm \epsfbox{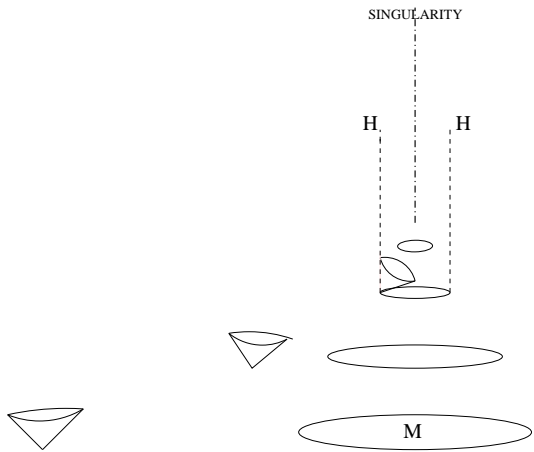}
\caption{Black hole spacetime emerging from spherically symmetric
collapse of a massive star M; ellipses denote successive sizes of the
collapsing star, and the null surface H is the event horizon}
\end{center}
\end{figure}

A sketch of the proof of Hawking's theorem may be given as follows
\cite{wald} :
Recall that the Raychaudhuri equation \cite{akr}
\begin{eqnarray}
l^a \nabla_a \theta \equiv {d\theta \over d\tau} ~=~
=~-\frac13~\theta^2~-~\sigma^2~+~\omega^2~-~R_{ab}~l^a~l^b~~
\label{rce}
\end{eqnarray}
relates the evolution of the (volume) expansion $\theta(\tau)$ of
(timelike and null) geodesics, to the `shear' $\sigma$ which
quantifies relative stretching of geodesics in a congruence, the
rotation or vorticity $\omega$ which exhibits how the geodesics twist
around  each other as they evolve and the spacetime curvature
responsible for geodesic deviation.

Consider any null surface generated by null geodesics which we assume
complete. A fixed time slice of such a surface is a spacelike 2-sphere
of area $A$. The Raychaudhuri equation (\ref{rce}) now implies that
\begin{eqnarray}
\theta^{-1}(\tau)~\geq~ \frac12 \tau~+~ \theta^{-1}(0)
\end{eqnarray}
The expansion $\theta = d \log A /d\tau$. Assume $\theta(0) > 0~$ in
contrast to the case for studying singularities; the area $A$ then
increases locally. Consider fixed time slices $S_1, S_2$ of event
horizon at $\tau_1,\tau_2 > \tau_1$. The number of null geodesic
generators  intersecting $S_2$ is at least the same or bigger than
those  intersecting $S_1$. At any instant, $A~\propto~no.~of~{\rm
Null~ geodesic~ generators}$; it follows then that $\Rightarrow~A_2
\geq A_1$ which is what the theorem states.

Hawking's theorem brings to mind the law of entropy increase (for
isolated systems) that follows from the second law of
thermodynamics. Yet, for a classical black hole spacetime, the
complete absence of `microstates' involving atoms, electrons, photons
or other discrete entities, stymies pushing the analogy very far. The
situation turns curiouser because of other `laws of (stationary) black
hole mechanics' derived on the basis of general relativity
\cite{wald}, wherein the geometrical quantity known as surface gravity
($\kappa$)  appears as an analogue of temperature.

This thermodynamic analogy of the theorems on stationary black hole
mechanics, especially, Hawking's theorem, led Bekenstein \cite{bek} to
{\it postulate} that {\it black holes must have an entropy
proportional to the area of their horizons}
\begin{eqnarray}
S_{bh}~=~\zeta~k_B~{{A} \over l_P^2} ~. \label{bek}
\end{eqnarray}
Here, $\zeta$ is a dimensionless constant of $O(1)$, and $l_P$  is the
Planck length $l_P \equiv (G \hbar /c^3)^{1/2} \sim 10^{-33}$ cm,
characterizing the length scale at which spacetime can no longer be
thought of in terms of classical Riemannian geometry. It follows that
black hole {\it thermodynamics} must have to do with {\it quantum},
rather than classical, general relativity. The microstates from which
$S_{bh}$ originates must then correspond to the states of the quantum
spacetime associated with the black hole, especially the event horizon.
    
Similarly, black hole temperature also has quantum origins, being
associated with the mysterious Hawking radiation \cite{hawr}. Every
generic black hole radiates particles in a thermal (black body)
spectrum, at a temperature $T_{bh} = \hbar \kappa_{horizon}$, a
surprising result, if one remembers that the surface gravity $\kappa$
is purely a geometric quantity. Hawking radiation thus corresponds to
{\it transformation} of quantum spacetime states describing the black
hole, into thermal particle states. In a sense, it is an inversion of
gravitational collapse whereby particle states collapse into spacetime
states. The extra feature here is the thermalization of the particle
states produced through Hawking radiation, which eventually leads to
the Information Loss conundrum.

The issue we address here is that of thermal stability of generic
radiant black holes in the heat bath made up of their own radiation.
The asymptotically flat Schwarzschild spacetime is well-known
\cite{hawp} to have a thermal {\it instability}: the Hawking
temperature for a Schwarzschild black hole of mass $M$ is given by $T
\sim 1/M$ which implies that the specific heat $C \equiv \partial
M/\partial T <0 ~!$ The instability is attributed, within a  standard
canonical  ensemble approach, to the superexponential growth of the
density of states  $\rho(M) \sim \exp M^2 $ which results in the
canonical partition function diverging for large $M$.

The problems with an approach based on an equilibrium canonical
ensemble do not exist, at least for isolated spherically  symmetric
black holes, formulated as {\it isolated horizons} \cite{al1} of fixed
horizon  area; these can be consistently described in terms of an
equilibrium {\it microcanonical} ensemble with fixed ${A}$ (and hence
disallowing thermal fluctuations of the energy $M$).  For ${A} >>
l_{Planck}^2$, it has been shown using Loop Quantum Gravity
\cite{abck}, that all spherically symmetric four dimensional isolated
horizons possess a microcanonical entropy obeying the
Bekenstein-Hawking Area Law (BHAL) \cite{bek},\cite{hawr}. Further,
the microcanonical entropy has corrections to the BHAL due to quantum
spacetime fluctuations at fixed horizon area. These arise, in the
limit of large ${A}$, as an  infinite series in inverse powers of
horizon area  beginning with a term logarithmic in the area \cite{km},
with  completely finite coefficients,
\begin{eqnarray}
S_{MC} =S_{BH} -\frac32 \log S_{BH} + const.+ {\cal O}(S_{BH}^{-1})
~.\label{smc}
\end{eqnarray}
where $S_{BH} \equiv {A}/4 l_{Planck}^2$

On the other hand, asymptotically anti-de Sitter (adS) black holes
with spherical symmetry are known \cite{hawp} to be describable in
terms of an equilibrium canonical ensemble, so long as the
cosmological constant is large in magnitude. For this range of black
hole parameters, to leading order in ${A}$ the canonical entropy obeys
the BHAL. As the magnitude of the cosmological constant is reduced,
one approaches the so-called Hawking-Page phase transition to a
`phase' which exhibits the same thermal instability as mentioned above.

In this article, we focus on the following
\begin{itemize}
\item Is an understanding of the foregoing features of  black hole
entropy and thermal stability on some sort of a `unified'  basis
possible ? We shall argue, following \cite{cm1}-\cite{cm5} that it is
indeed so, at least in the case of non-rotating black holes.
\item In addition to corrections (to the area law) due to fixed  area
quantum spacetime fluctuations computed using a microcanonical
approach, can one compute corrections due to {\it  thermal}
fluctuations of horizon area within the canonical ensemble  ? Once
again, the answer is in the affirmative. The  result found in
\cite{cm1}-\cite{cm5}, at least for the leading  $\log area$  corrections,
turns out  to be {\it universal} in the sense that, just like the
BHAL, it  holds for all black holes independent of their parameters.
\end{itemize}

\section{Canonical partition function : holography ?}

Following \cite{cm1}, we start with the canonical partition function
in the quantum case
\begin{eqnarray}
Z_C(\beta)~=~Tr~\exp -\beta \hat{H} ~.
\label{qpf}
\end{eqnarray}
Recall that in classical general relativity in the Hamiltonian
formulation, the bulk Hamiltonian is a first class constraint, so that
the entire Hamiltonian consists of the boundary contribution $H_S$ on
the constraint surface. In the quantum domain, the Hamiltonian
operator can be written as
\begin{eqnarray}
\hat{H}~=~\hat{H}_V~+~\hat{H}_S~,
\label{deco}
\end{eqnarray}
with the subscripts $V$ and $S$ signifying bulk and boundary terms
respectively. The Hamiltonian constraint is then implemented by
requiring
\begin{eqnarray}
\hat{H}_V~|\psi \rangle_V~=~0~
\label{hcon}
\end{eqnarray}
for every physical state $|\psi \rangle_V$ in the bulk. Choose as
basis for the Hamiltonian in (\ref{deco}) the states $|\psi  \rangle_V
\otimes |\chi\rangle_S$. This implies that the  partition function may
be factorized as
\begin{eqnarray}
Z_{C}~&\equiv&~ Tr \exp -\beta {\hat H} \nonumber \\
&=&~\underbrace{dim~{\cal  H}_V}_{{indep.~of~\beta.}}~\underbrace{Tr_S
\exp -\beta {\hat H}_S}_{{boundary}} \label{facto}
\end{eqnarray}
Thus, the relevance of the bulk physics seems rather limited due to
the constraint (\ref{hcon}). The partition function further reduces  to
\begin{eqnarray}
Z_C(\beta)~=~dim~{\cal H}_V~Z_S(\beta)~,
\label{redp}
\end{eqnarray}
where ${\cal H}_V$ is the space of bulk states $|\psi\rangle$ and
$Z_S$ is the `boundary' partition function given by
\begin{eqnarray}
Z_S(\beta)~=~Tr_S~\exp -\beta \hat{H}_S~.
\label{zs}
\end{eqnarray}

Since we are considering situations where, in addition to the boundary
at asymptopia, there is also an inner boundary at the black hole
horizon, quantum fluctuations of this boundary lead to black hole
thermodynamics. The factorization in eq.(\ref{redp}) manifests in the
canonical entropy as the appearance of an additive constant
proportional to $dim~{\cal H}_V$. Since thermodynamic entropy is
defined only upto an additive constant, we may argue that the bulk
states do not play any role in black hole thermodynamics. This may be
thought of as the  origin of a weaker version of the holographic
hypothesis \cite{thf}.

For our purpose, it is more convenient to rewrite (\ref{zs}) as
\begin{eqnarray}
Z_C(\beta)~~=~~\sum_{n \in {\cal Z}} \underbrace{g\left (E_S({A}(n))
\right)}_{{degeneracy}}~\exp  -\beta E_S({A}(n))~, \label{zss}
\end{eqnarray}
where, we have made the assumptions that (a) the energy is a function
of the area of the horizon ${A}$ and (b) this area is quantized.  The
first assumption (a) basically originates in the idea in the last
paragraph of that black hole thermodynamics ensues solely from the
boundary states whose energy ought to be a function of some property
of the boundary like area. The second assumption (b) is actually
explicitly provable in theories like NCQGR as we now briefly digress
to explain.

\section{Spin network basis in NCQGR}

The basic canonical degrees of freedom in NCQGR are holonomies of a
distributional $SU(2)$ connection and fluxes of the densitized triad
conjugate to this connection. The Gauss law (local $SU(2)$ invariance)
and momentum (spatial diffeomorphism) constraints are realized as
self-adjoint operators constructed out of these variables. States
annihilated by these constraint operators span the kinematical Hilbert
space. Particularly convenient bases for this kinematical Hilbert
space are the spin network bases. In any of these bases, a (`spinet')
state is described in terms of {\em links} $l_1, \dots, l_n$ carrying
spins ($SU(2)$ irreducible representations) $j_1, \dots j_n$ and {\it
vertices} carrying invariant $SU(2)$ tensors (`intertwiners'). A
particularly important property of such bases is that geometrical
observables like area operator are diagonal in this basis with {\em
discrete spectrum}. An internal boundary of a spacetime like a horizon
appears in this kinematical description as a punctured ${\cal S}^2$
with each puncture having a deficit angle $\theta=\theta(j_i),
i=1,\dots,p$, as shown in Fig.1.

\begin{figure}
\epsfxsize=60mm \centerline{\epsfbox{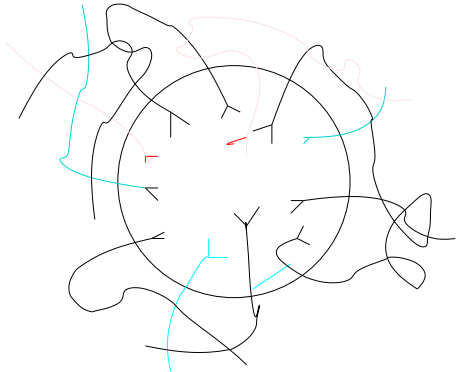}}
\caption{Internal boundary (horizon) pierced by spinet  links}
\label{fig:spinet}
\end{figure}

For macroscopically large boundary areas ${A} >> l_{Planck}^2$,  the
area spectrum is dominated by $j_i=1/2, \forall i=1, \dots,p,~ p
>>1$. This is the situation when the deficit angles at each puncture
takes its smallest nontrivial value, so that a classical horizon
emerges. That implies that
\begin{eqnarray}
{A}(p) ~\sim ~p~ l_{Planck}^2~,~p~>>~1  ~. \label{area}
\end{eqnarray}
This completes our digression on NCQGR.

\section{Fluctuation effects on canonical entropy}

We now move on to discuss the effect of inclusion of Gaussian thermal
fluctuations of the horizon area. The canonical entropy is expected to
receive additional corrections due to such fluctuations over and above
those due to quantum spacetime fluctuations already included in the
microcanoncal entropy. Going back to eq. (\ref{zss}), we can now
rewrite the partition  function as an integral, using the Poisson
resummation formula
\begin{eqnarray}
\sum_{n=-\infty}^{\infty} f(n)~=~\sum_{m=-\infty}^{\infty}
\int_{-\infty}^{\infty} dx ~\exp (-2\pi i mx)~f(x)~. \label{poi}
\end{eqnarray}  
For macroscopically large horizon areas ${A}(p)$, $x >> 1$,  so that
the summation  on the {\em rhs} of (\ref{poi}) is dominated  by  the
contribution of the $m=0$ term. In this approximation, we have
\begin{eqnarray}
Z_C &~\simeq~& \int_{-\infty}^{\infty} dx~g(E(A(x)))~\exp -\beta
E(A(x)) \nonumber \\ &=&~\int dE~\exp [S_{MC}(E)~-~\log|{dE \over
dx}|~-~\beta E]
\label{appr}
\end{eqnarray}
where $S_{MC} \equiv \log g(E)$ is the microcanonical entropy.

Now, in equilibrium statistical mechanics, there is an inherent
ambiguity in the definition of the microcanonical entropy, since it
may also be defined as ${\tilde S}_{MC} \equiv \log \rho(E)$ where
$\rho(E)$ is the density of states. The relation between these two
definitions involves the `Jacobian' factor $|dE /dx|^{-1}$
\begin{eqnarray} 
{\tilde S}_{MC}~=~S_{MC}~-~\log|{dE \over dx}| ~.\label{amb}
\end{eqnarray}
Clearly, this ambiguity is irrelevant if all one is interested in is
the leading order BHAL. However, if one is interested in logarithmic
corrections to BHAL as we are, this difference is crucial and must be
taken into account.

We next proceed to evaluate the partition function in eq.
(\ref{appr}) using the saddle point approximation around the point
$E=M$ where $M$ is to be identified with the (classical) mass of the
boundary (horizon). Integrating over the Gaussian fluctuations around
the saddle point, and dropping higher order terms, we get,
\begin{eqnarray}
Z_C &\simeq& \exp {\{ S_{MC}(M) - \beta M - \log|{dE \over dx}|_{E=M}]
\}} \nonumber \\ & \cdot & \left [{\pi \over -S_{MC}''(M)}
\right]^{1/2}~.
\label{spa}
\end{eqnarray}
Using $S_C = \log Z_C + \beta M$, we obtain for the canonical  entropy
$S_C$
\begin{eqnarray}
S_C~&=&~S_{MC}(M)~\underbrace{-~\frac12
\log(\Delta)}_{{\delta_{th}S_C}}~, \label{scan}
\end{eqnarray}
where,
\begin{eqnarray}
\Delta~\equiv~[{A}'(x)]^2~\left[ S_{MC}'({A}) ~{M''({A}) \over
M'({A})}~-~S_{MC}''({A}) \right]~.
\label{delta}
\end{eqnarray}
Thus, the canonical entropy is expressed in terms of the
microcanonical entropy for an average large horizon area, and the mass
which is also a function of the area. Clearly, stable  equilibrium
ensues so long as $\Delta > 0$.

Additional support for this condition can be gleaned by considering
the thermal capacity of the system, using the standard relation
\begin{eqnarray}
C({A}) \equiv {dM \over dT} = {M'({A}) \over T'({A})},
\label{hea}
\end{eqnarray}
with $T$ being derived from the microcanonical entropy $S_{MC}({A})$,
and hence a function of ${A}$. One obtains for the heat capacity the
relation
\begin{eqnarray}
C({A})~=~\left[{M'({A}) \over T({A}) {A}'(x)} \right]^2~\Delta^{-1} ~,
\label{hc}
\end{eqnarray}
so that $C > 0$ if only if $\Delta > 0$. Since the positivity of the
heat capacity is certainly a necessary condition for stable thermal
equilibrium, it is gratifying that an identical criterion emerges for
$\Delta$ as found from the canonical entropy (\ref{scan}).

Using now eq, (\ref{delta}) for the expression for $\Delta$, the
criterion for thermal stability of non-rotating macroscopic black
holes is then easily seen to be
\begin{eqnarray}
M({A}) ~>~ S_{MC}({A}) ~
\label{crit}
\end{eqnarray}
as already mentioned in the summary. We have been using units in which
$G=\hbar=c=k_B=1$. If we revert back to units where these constants
are not set to unity, the lower bound eq. (\ref{crit}) can be
re-expressed as
\begin{eqnarray}
M({A}) ~>~ \left({\hbar c \over G k_B^2} \right)^{1/2} S_{MC}({A})
~. \label{crit2}
\end{eqnarray}
We remind the reader that in contrast to semiclassical approaches
based on specific properties of classical metrics, our approach
incorporates crucially the microcanonical entropy generated by quantum
spacetime fluctuations that leave the horizon area constant. Apart
from the plausible assumption of the black hole mass being dependent
only on the horizon area, no other assumption has been made to arrive
at the result. Even so, it subsumes most results based on the
semiclassical approach.

As a byproduct of the above analysis, the canonical entropy for stable
black holes can be expressed in terms of the Bekenstein-Hawking
entropy $S_{BH}$ as
\begin{eqnarray}
S_C~&=&~S_{BH} -\frac12 (\xi- 1) \log S_{BH} \nonumber \\ ~&-&~\frac12
\log \left[ {S'_{MC}({A}) ~ M''({A}) \over S_{MC}''({A}) M'({A})}
\right] ~.\label{sc}
\end{eqnarray}  
For any smooth $M({A})$, one can truncate its power series expansion
in ${A}$ at some large order and show that the quantity in square
brackets in eq. (\ref{sc}) does not contribute to the $\log (area)$
term, so that
\begin{eqnarray}
S_C~=~S_{BH}~-\frac12~(\xi-1)~\log S_{BH}~+~const.~+~O(S_{BH}^{-1}) ~,
\label{scano}
\end{eqnarray}
where, $\xi =3$ in eq. (\ref{smc}). Note that this is the result for
an isolated horizon described by an $SU(2)$ Chern Simons theory. For a
$U(1)$ Chern Simons theory, $\xi=1$ \cite{dkm},\cite{cm3}. The interplay between
constant area quantum spacetime fluctuations and thermal fluctuations
is obvious in the coefficient of the $log(area)$ term where the
contribution due to each appears with a specific sign. It is not
surprising that the thermal fluctuation contribution increases the
canonical entropy. The cancellation occurring for horizons on which a
residual $U(1)$ subgroup of $SU(2)$ survives, because of additional
gauge fixing by the boundary conditions describing an isolated horizon
\cite{al1}, may indicate a possible non-renormalization theorem,
although no special symmetry like supersymmetry has been employed
anywhere above. It is thus generic for all non-rotating black holes,
including those with electric or dilatonic charge.

While so far we have restricted our attention to thermal fluctuations
of area due to energy fluctuations alone, the stability criterion
(\ref{crit}) can be shown to hold when in addition thermal
fluctuations of electric charge are incorporated within a grand
canonical ensemble \cite{cm4}. As in \cite{cm2}, we assume that energy
spectrum is a function of the discrete area spectrum (well-known in
LQG \cite{al2}) and a discrete charge spectrum. The charge spectrum is
of course equally spaced in general;   for large macroscopic black
holes the area spectrum is equally spaced as well.

In a basis in which both the area and charge operators are
simultaneously diagonal, the grand canonical partition function can be
expressed as
\begin{eqnarray}
Z_G~=~ \sum_{m,n}~g(m,n)~\exp - \beta \left[~ E (A_m,Q_n)~-~\Phi~Q_n
\right] ~, \label{gpf5}
\end{eqnarray}
where, $g(m,n)$ is the degeneracy corresponding to the area eigenvalue
$A_m$ and charge eigenvalue $Q_n$.  Using a generalization of the
Poisson resummation formula
\begin{eqnarray}
\sum_{m,n}~f(m,n)~=~\sum_{k,l}~\int~dx~dy ~\exp ~\{-i(kx +
ly)\}~f(x,y) ~\label{poic}
\end{eqnarray}
and assuming that the partition sum is dominated by the large
eigenvalues $A_m~,Q_n$, it can be expressed as a double integral
\begin{eqnarray}
Z_G~=~\int ~dx~dy~\exp ~-~\beta ~\{ E(A(x),Q(y))~-~\Phi~Q(y)
\}~g(A(x),Q(y))~. \label{gpf6}
\end{eqnarray}
Note that the transition from the discrete sum to the integral for
$Z_G$ requires only that the dominant eigenvalues are large compared
to the fundamental units of discreteness which for the area is the
Planck area and for the charge is the electronic charge. These
conditions are of course fulfilled for all astrophysical black holes.

Changing variables in eq. (\ref{gpf6}) form $x,y$ to $E,Q$
\begin{eqnarray}
Z_G~ &=& ~\int dE~dQ~{\cal J}(E,Q)~g(E,Q)~\exp~\{- \beta (E- \Phi  Q)
\}~ \nonumber \\ &=&~\int dE~dQ~\rho(E,Q)~\exp~\{ - \beta (E- \Phi Q)
\} ~,
\label{gpf7}
\end{eqnarray}
where, the Jacobian ${\cal J}~=~\left|~ E_{,x} ~\right|^{-1} ~  \left|
~Q_{,y}~ \right|^{-1} $, and $\rho = {\cal J}(E,Q)~g(E,Q)$ is  the
density of states. Employing the saddle point approximation and  using
eq. (\ref{sg}) one obtains
\begin{eqnarray}
S_G(M,Q_0)~=~{\tilde S}_{MC}(M,Q_0)~-~\frac12~\log \Delta ~+~const. ~
\label{sg22}
\end{eqnarray}
where, using $S_{MC} = \log \rho = {\tilde S}_{MC} + \log {\cal J}$ we
have defined
\begin{eqnarray}
\Delta ~\equiv~ \det \Omega~{\cal J}^2~& = &~\det
\Omega~(E_{,x})^2~(Q_{,y})^2|_{M,Q_0} \nonumber \\ & = &~ \det
\Omega~(A_{,E})^2 ~(A_{,x})^2~ (Q_{,y})^2|_{M,Q_0} ~,
\label{del}
\end{eqnarray}
where, the Hessian matrix
\begin{eqnarray}
\Omega ~=~ \pmatrix{ S_{MC,EE} &  S_{MC,EQ} \cr S_{MC,EQ} &  S_{MC,QQ}
}_{M,Q_0} ~.  \label{omeg}
\end{eqnarray}

Since the microcanonical entropy is known to be only a function of the
horizon area even for {\it charged} non-rotating black holes
\cite{al1,km}, one can express $\det \Omega$ as
\begin{eqnarray}
\det \Omega ~ & = & ~(S_{MC,A})^2~\left[ (A_{,EE})~(A_{,QQ}) -
(A_{,EQ})^2 \right] |_{M,Q_0} \nonumber \\ &+&
S_{MC,AA}~S_{MC,A}~\left[ (A_{,E})^2~A_{,QQ} + (A_{,Q})^2~A_{,EE}  -
2~A_{,E}~A_{,Q}~A_{,EQ} \right] |_{M,Q_0} ~\label{omeg2}
\end{eqnarray}

The necessary and sufficient conditions for thermal stability are
\begin{eqnarray}
Tr \Omega ~&=& ~ S_{MC,EE} \left|_{M,Q_0}~+~S_{MC,QQ} \right|_{M,Q_0}~
< ~ 0 \nonumber \\ \det \Omega ~&=& ~ \left[
~S_{MC,EE}~S_{MC,QQ}~-~S_{MC,EQ}^2 ~ \right]_{M,Q_0}~>~ 0 ~
\label{posd}
\end{eqnarray}
which necessarily imply
\begin{eqnarray}
S_{MC,EE}|_{M,Q_0}~<~0~,~ \mbox{and}~  S_{MC,QQ}|_{M,Q_0}~<~0~.
\label{psd}
\end{eqnarray}
Note that while these conditions together imply the first of the
necessary and sufficient conditions (\ref{posd}) for stability of
$\Omega$, they are not sufficient to guarantee the second one.

Using the microcanonical relations for temperature and potential, we
may expess the necessary conditions for stability in terms of the heat
capacity $C_Q \equiv (\partial E/\partial T)_Q$ and the capacitance $C
\equiv (\partial Q / \partial \Phi)_E$ in the following way
\begin{eqnarray}
C_Q~~ > ~~ 0~~ \mbox{and}~~ C~\Phi~~ < ~~ T~\left( \partial Q \over
\partial T \right)_E ~.
\label{capa}
\end{eqnarray}
The more stringent necessary and sufficient conditions can also be
similarly expressed in terms of $C_Q$ and $C$.

The grand partition function, evaluated in the saddle point
approximation, can now be substituted in the standard thermodynamic
relation in the presence of a chemical (electrostatic) potential
\begin{eqnarray}
S_G~ = ~ \beta~M~-~\beta~Q~\Phi~+~ \log Z_G ~, \label{sg}
\end{eqnarray}
so as to yield the grand canonical entropy
\begin{eqnarray}
S_G~~=~~ S_{MC}~-~\frac12~ \log ~\det ~\Omega ~+~const. ~
\label{gentro}
\end{eqnarray}

We now make use of eq. (\ref{smc}) to observe that both $S_{{MC},A}$
and $S_{{MC},AA}$ are positive definite for macroscopically large
areas. The necessary and sufficient conditions for thermal stability,
under both energy and charge fluctuations, can now be transcribed into
the three simple inequalities
\begin{eqnarray}
{A_{MM} \over A_{M}^2} ~+~ { S_{{MC},AA} \over S_{{MC},A} }~ & < &~ 0
\label{mas} \cr
{A_{QQ} \over A_{Q}^2} ~+~ { S_{{MC},AA} \over S_{{MC},A} }~ & < &~ 0\
\label{cha} \cr
{A_{MM} \over A_M^2}~{A_{QQ} \over A_Q^2} ~& > &~ \left( { S_{{MC},AA}
\over S_{{MC},A} } \right)^2  ~,
\label{deto}
\end{eqnarray} 
where, we have used the notation $A_{,EE}|_{M,Q_0} \equiv A_{MM}$
etc. We have assumed of course that the horizon area $A =
A(M,Q)$. Since we are identifying the black hole mass $M$ with the
mass defined for the isolated non-radiating non-accreting horizon, we
can recall our earlier assumption that $M = M(A)$, so that $A =
A(M(A),Q)$. In other words, the mass associated with the isolated
horizon ought to relate to the horizon area just as the bulk
Hamiltonian relates to the volume operator (as shown in
ref. \cite{thie}). The quantum mass spectrum thus should be related to
the area spectrum. On the other hand, electric charge has no such
geometric origin, and is independent of the area, as far as one can
make out. This implies that
\begin{eqnarray}
A_M ~M'(A) ~&=&~1 \nonumber \\ A_{MM}~ M'^2(A)~& + &~ A_M~M''(A)~ = ~
0~. \label{deja}
\end{eqnarray}
Substituting eq. (\ref{deja}) into the inequality (\ref{mas}), the
{\bf stability criterion eq. (\ref{crit}) emerges once again}. It is
not difficult to see that this, when substituted back into
(\ref{deto}), merely reproduces (\ref{cha}), thereby establishing
consistency between the inequalities.

Is it possible to derive as a bonus, as in the case of non-fluctuating
charge, a general formula for the thermal fluctuation correction to
the canonical entropy, logarithmic in the horizon area ? We believe it
is indeed the case, but do not include that discussion here.

Can this criterion continue to hold if the black hole is characterized
by a number of $U(1)$ charges $Q_1, \dots, Q_{n-1}$, all independently
quantized ?  This situation typically ensues in black holes arising in
the low energy supergravity limit of string theories. The matrix
$\Omega$ is then an $n \times n$ matrix, and the analogues of eq.s
(\ref{mas})-(\ref{deto}) now become more complicated, involving sums
of products of derivatives of $S_{MC}$. Despite this, so long as one
stays away from extremality, it is not inconceiveable that the mass
alone decides on the stability issue.

\section{Concluding remarks}

The laws of black hole mechanics derived from the Raychaudhuri
equation have inevitably led to worldwide attempts to seek quantum
formulations of general relativity. A reasonably satisfactory
understanding of black hole entropy has been achieved : the entropy of
radiant black holes is completely described in terms of the
microcanonical entropy of isolated horizons which, in turn, is more or
less entirely understood within the scheme of Loop Quantum Gravity, at
least for the cases without rotation. Inclusion of rotation into the
LQG approach to microcanonical entropy remains now foremost on the
agenda, since, like the mass of the isolated horizon, and unlike
$U(1)$ charges, the angular momentum is also likely to be determined
by the horizon area in the isolated case.

\vglue .3cm

\noindent {\bf Acknowledgement} I thank A. Chatterjee for several
illuminating discussions.

\end{document}